\documentclass[12pt]{article}

\usepackage{graphicx}
\usepackage{rangecite}
\usepackage{ifpdf}

\def\be{\begin{equation}}
\def\ee{\end{equation}}
\def\bear{\begin{eqnarray}}
\def\eear{\end{eqnarray}}
\def\nn{\nonumber}

\def\half{{{1\over 2}}}

\newcommand\tr{{\mbox{tr}}}          

\def\a{{\alpha}}
\def\b{{\beta}}

\def\Om{{\Omega}}

\begin{document}


\begin{titlepage}
\vskip 1in
\begin{center}
{\Large{Scattering in the adjoint sector of the $c=1$ Matrix Model}}
\vskip 0.5in
{Joanna L. Karczmarek}
\vskip 0.3in
{\it 
Department of Physics and Astronomy\\
University of British Columbia
Vancouver, Canada}
\end{center}
\vskip 0.5in
\begin{abstract}
Closed string tachyon emission from a traveling long string in Liouville
string theory is studied.  The exact collective field Hamiltonian in the
adjoint sector of the c=1 matrix model is computed to capture the
interaction between the tip of the long string and the closed string tachyon
field.  The amplitude for emission of a single tachyon quantum is obtained
in a closed form using the chiral formalism.
\end{abstract}
\end{titlepage}



\section{Introduction}

As is very well known, the singlet sector of  Matrix
Quantum Mechanics of a hermitian matrix 
in an inverted harmonic oscillator potential 
(known as the c=1 matrix model)
is equivalent to Liouville string theory in two spacetime dimensions.
Liouville conformal field theory amplitudes on closed surfaces
are believed to be reproduced by the scattering amplitudes in the
MQM (with the genus expansion corresponding to semi-classical
expansion of MQM).  In contrast, the spacetime and worldsheet
interpretation of the non-singlet sectors of the same
Matrix Quantum Mechanics is less clear, especially under
Lorentzian time signature\footnote{Somewhat more is known about the theory in
Euclidean time \cite{Kazakov:2000pm}; the intriguing results in 
\cite{Kazakov:2000pm}
are one motivation for exploring the Lorentzian signature
further.}.

In \cite{Maldacena:2005hi} it was proposed that
the adjoint sector corresponds to Liouville string theory
in the presence of a single long string.  
In the spacetime picture, the ends of the long string are at infinity, 
in the asymptotically free region
away from the Liouville wall.  The string itself is folded in
half, and the place where it folds, the `tip', moves
under combined effects of string tension and stored
kinetic energy.  The tip moves nearly with the speed of light
towards the Liouville wall until the long string is fully stretched, 
and then  `snaps' back to the free region.  Depending on
its kinetic energy, it might or might not travel far enough into
the bulk of  spacetime to `hit' the Liouville wall.

The worldsheet interpretation of the adjoint sector requires
addition of an open string boundary with two
boundary operator insertions corresponding to the incoming and 
outgoing state of the long string.  To keep the ends of the long
string at infinity, the boundary can be taken to be an 
FZZT brane with a very large boundary cosmological constant 
(as compared to the bulk cosmological constant).
The energy of the boundary operators must be large enough to
allow for the string to be thought of as `long' (see 
\cite{Maldacena:2005hi}
for details of the limits involved).

This picture allows one to use the worldsheet point of view 
to study scattering of such long  strings.  This was
done for a single long string in \cite{Maldacena:2005hi}
(the scattering amplitude is simply the disk
boundary two point function in the appropriate limit). 
and for multiple long strings in \cite{Bourgine:2007ua}.
Worldsheet  results were found to agree with matrix model results in 
\cite{Maldacena:2005hi,Fidkowski:2005ck,Donos:2005vm,Kostov:2006dy,Bourgine:2007ua}.

The main purpose of the present paper is to extend
previous work on scattering of long strings to include backreaction.
This should allow us to further test the long string proposal,
as well as explore the physics of long strings.
In particular, we will be studying emission of closed string
tachyons, the only degrees of freedom in
Liouville string theory which can travel to null infinity.
Emission of a single tachyon quantum is shown in Figure 1.
This emission can be obtained
from a disk diagram with two boundary insertions (for the 
incoming and outgoing long string) and one bulk tachyon insertion.
In the present paper, we will focus on computing this
amplitude in the matrix model, and leave the conformal
theory calculation to future work.
The situation pictured in Figure 1(b) is probably easier
from the worldsheet point of view, since the bulk cosmological
constant can be set to zero.

\begin{figure}
\ifpdf
\includegraphics[scale=0.65]{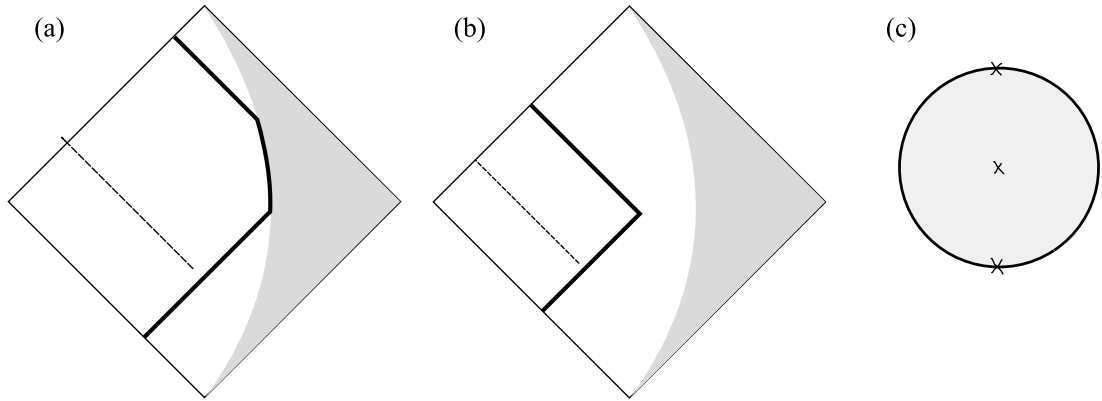}
\else
\includegraphics[scale=0.65]{long-string-emits-tachyon.eps}
\fi
\caption{A long string emitting a tachyon. In the Penrose diagrams in
(a) and (b), the tip of the long string is drawn as a heavy line,
the emitted tachyon is a dashed line, and the shaded region is
the region `beyond the Liouville wall' where strings cannot propagate.
{\bf (a)} A high energy long string. 
{\bf (b)} A low energy long string, which does not reach the Liouville wall.
{\bf (c)} The corresponding three point disk amplitude. The 
boundary of the disk must have a large cosmological constant.
\label{f1}}
\end{figure}


In Matrix Quantum Mechanics, restricting to the adjoint sector (as
opposed to the singlet sector) corresponds to introducing an 
interacting impurity into the theory of free fermionic matrix eigenvalues.
This impurity corresponds to the tip of the long string,
and we would like to study its interactions with the
background eigenvalue density.

In section \ref{s1},
we review quantization of Matrix Quantum Mechanics
in any sector, including the adjoint.
Following this, we study interaction between the 
adjoint impurity  and the rest of the eigenvalues in two ways.

In sections \ref{s2}-\ref{s4}, we use standard methodology 
\cite{Donos:2005vm,Das:1990kaa,Jevicki:1979mb}
to derive the collective field Hamiltonian describing the interaction 
between the impurity and the density of eigenvalues.  Our
result at the end of section \ref{s2} 
extends that obtained in \cite{Donos:2005vm},
where only quadratic terms in the Hamiltonian were computed. 
In section \ref{s3}, we
derive the same result using second quantized fermionic description,
and bosonization.
In section \ref{s4}, we discus the form of the collective field
Hamiltonian and some of the surrounding technical issues.  
The results on the collective field Hamiltonian in sections
\ref{s2} and \ref{s3} are applicable to any potential.

In section \ref{s5}, we
derive the amplitude for emission of a single tachyon quantum
during long string scattering.  We use the chiral formalism \cite{Alexandrov:2002fh},
adopted for computations involving long strings in
\cite{Kostov:2006dy} and used to describe
scattering of multiple long strings in \cite{Bourgine:2007ua}.
Our approach here is suitable only for the inverted harmonic
oscillator potential. The final result of this calculation,  
given in equation (\ref{eqn:answerb}),
should reproduce the disk amplitude in Figure \ref{f1}(c)
once all the appropriate leg-pole factors are taken into
the account.

\section{Quantization of the matrix model}
\label{s1}

In this section, for the sake of completeness, we review
the quantization of an  ungauged one-matrix model with an
arbitrary potential $V$.  The classical Lagrangian is
\be
\label{L}
{\cal L} (M) = \half \tr (\dot M)^2 - \tr V(M)~,
\ee
where $M$ is a hermitian $N\times N$ matrix,
while the quantum Hamiltonian can be written simply as
\be
H = - \half {\partial \over \partial M_{ab}} {\partial \over \partial M_{ba}}
+ \tr V(M)~,
\label{eqn:H}
\ee
when acting on a wavefunction $\Psi(M)$.  
The lower case Roman indices are matrix indices and run from $1$ to $N$.
The kinetic term is just
that of $N^2$ free particles, but the potential couples these in
a complicated matter.  To take advantage of the (global) $SU(N)$ 
gauge symmetry present in this problem:
\be
SU(N):~~M~\rightarrow U M U^\dagger~,
\ee
it is helpful to decompose $M$ into a diagonal part $\Lambda$ and 
an angular part $\Omega \in SU(N)$, $M= \Omega \Lambda \Omega^\dagger$, 
and then study the wavefunctions in each irrep of $SU(N)$,
keeping in mind that the decomposition into $\Lambda$ and 
$\Omega$ is not unique.
We will follow the approach of \cite{Marchesini:1979yq}.
Consider a basis of wavefunctions $\Psi_\a$ spanning
an irreducible representation of $SU(N)$
\be
\Psi_\a(U M U^\dagger) = U_{\a\b} \Psi_\b(M)~.
\ee
The Greek indices are $SU(N)$ representation indices.
The Hamiltonian does not couple different irreducible representations
to each other.

Consider a point where $M$ is diagonal, $M=\Lambda$.  Then, for
an action of a diagonal $SU(N)$ matrix $D$ we have
\be
\Psi_\a(\Lambda) = \Psi_\a(D \Lambda D^\dagger) = 
D_{\a\b} \Psi_\b(\Lambda)~.
\ee
Therefore, $\Psi_\a(\Lambda)$ must be a zero-weight vector in the
irrep under consideration (transforms trivially under the cartan
subgroup).  We need then to restrict our attention to irreps with
zero-weight vectors.  If we denote by $v^{(k)}_\a$ a basis for 
the zero-weight subspace of the representation and by K its
dimension, we must have
\be
\Psi_\a(\Lambda) = \sum_{i=1}^K f^{(k)}(\Lambda) v^{(k)}_\a~,
\ee
and
\be
\Psi_\a(M = \Omega \Lambda \Omega^\dagger) = 
\sum_{k=1}^K f^{(k)} (\lambda_1, \lambda_2,..,\lambda_N) \Omega_{\a\b} v^{(k)}_\b~,
\label{eqn:psi}
\ee
where $\lambda_i$ are the eigenvalues of $M$.
The problem is therefore reduced to finding the $K$ `radial' 
functions $f^{(k)}(\lambda_i)$.

Notice that, by construction, this is well defined, since
the ambiguity in decomposition of $M$ into $\Omega$ and $\Lambda$
does not affect anything.  When $\Omega$ is multiplied on the
right by a diagonal matrix $D$, we have
\be
(\Omega D)_{\a\b} v^{(k)}_\b~ = (\Omega)_{\a\b} (D)_{\b\gamma} 
v^{(k)}_\gamma~ v^{(k)}_\b~ = (\Omega)_{\a\b} v^{(k)}_\b~.
\ee
Another possible ambiguity arises when the eigenvalues are permuted.
Given a permutation $\sigma \in S_N$, let $P$ be a permutation matrix
$P = \delta_{i, \sigma(i)}$.  $P$ is not necessarily in $SU(N)$, as
$\det(P) = (-1)^\sigma$, the sign of the permutation is not necessarily 1.
We define a matrix in $SU(N)$ as
\be
\tilde P = ((-1)^\sigma)^{1/N} P = \exp \left ({1 -(-1)^\sigma \over 2} 
{\pi i \over N} \right ) P
\label{eqn:Ptilde}
\ee
Then,
\be
(\Omega \tilde P)_{\a\b} v^{(k)}_\b~ = (\Omega)_{\a\b} \tilde P_{\b\gamma} 
v^{(k)}_\gamma = (\Omega)_{\a\b} \sum_{(k)} \tilde P_{(k)(l)} v^{(l)}_\b~,
\ee
where $\tilde P_{(k)(l)}$ is the modified permutation matrix in the zero-weight
space representation.
This means that there is an action on the indices (k) induced by P 
and we obtain a symmetry constraint on the f's
\be
f^{(k)}(\lambda_{\sigma(i)}) =\sum_{(k)} \tilde P_{(k)(l)} f^{(l)}(\lambda_i)~.
\label{eqn:permutations}
\ee

It is worth stressing that the origin of the condition 
(\ref{eqn:permutations}) is due 
solely to the non-uniqueness of the decomposition 
$M = \Omega \Lambda \Omega^\dagger$.

Naively, the action of $\sigma$ on the zero weight space given
by (\ref{eqn:permutations}) and (\ref{eqn:Ptilde}) is badly defined.  
In particular,
let $P_1$ and $P_2$ be odd permutations, and $P=P_1 P_2$. Then $P=\tilde P$ 
while
\be
\tilde P_1 \tilde P_2 = \exp \left ({2 \pi i \over N} \right ) P_1 P_2~.
\ee
In fact, the $\exp ({2 \pi i / N})$ factor does not cause a problem 
only because it acts trivially on the zero weight space (the
determinant is equal to 1).

Note that for $K=1$, the unique zero-weight vector is
not necessarily invariant under the action of $\sigma$.
The symmetric group $S_N$ has two one dimensional representations;
for the trivial representation,  $f(\lambda_i)$
is a totally symmetric function of the eigenvalues, and for
the alternating representation, $f(\lambda_i)$ must be totally
antisymmetric.  For example, for the totally symmetric
representation of $SU(N)$ with $kN$ boxes, we have $K=1$,
and the wavefunctions are completely symmetric
when $k$ is even, and completely antisymmetric when $k$ is odd.

The allowed representations are those which have at least one
zero-weight state.  This condition is equivalent to considering
only those representations which arise as in the
decomposition of tensor powers of the adjoint representation,
which in turn is equivalent to requesting that the number of boxes in 
the Young tableaux must be divisible by N (which allows the diagram 
to be drawn in the simplified boxes-and-antiboxes form \cite{Gross:1990md}).

To obtain the Schrodinger equation, 
we need to rewrite the Hamiltonian in terms of the new variables
$\Omega$ and $\lambda_i$.  This can be accomplished with the
following two formulas\footnote
{For completeness, here is a derivation, which is essentially
identical with the derivation of first order perturbation theory.  
Notice that the columns of $\Omega$ are eigenvectors of $M$ 
with eigenvalues $\lambda_i$.  Consider the equation 
$M |i\rangle = \lambda_i |i\rangle$, where $|i\rangle$ is the 
$i^{th}$ column.  Varying each side, and hitting the
equation with $\langle j|$ on the left, we obtain 
$\langle j | \delta M |i\rangle +
(\lambda_i - \lambda_j) \langle j |(\delta | i \rangle) = 
(\delta \lambda_i) \delta_{ij}$.  The two formulas follow
easily from the off-diagonal and diagonal part of this
equation respectively.}
\bear
{\partial \Omega_{ki} \over \partial M_{ab}} &=& \sum_{j\neq i}
{\Om^\dagger_{ja} \Om_{bi} \over \lambda_i-\lambda_j} \Om_{kj}~,
\label{eqn:dOm}
\\
{\partial \lambda_i \over \partial M_{ab}} &=& \Om^\dagger_{ia} \Om_{bi}~.
\label{eqn:dL}
\eear
A complex conjugate of formula (\ref{eqn:dOm}) is also useful
\be
{\partial \Om^\dagger_{ik} \over \partial M_{ba}} = \sum_{j\neq i}
{\Om_{aj} \Om^\dagger_{ib} \over \lambda_i-\lambda_j} \Om^\dagger_{jk}~.
\ee
Armed with these formulas, it is possible to explicitly write
the action of the Hamiltonian (\ref{eqn:H}) on the wavefunction
(\ref{eqn:psi}), once the angular dependence contained in
$\Omega_{\a\b}$ is known.  While explicit, this
calculation is quite cumbersome for even the simplest
representations.

A much more efficient way of obtaining the Schrodinger 
equation in a given representation is to start with the
Lagrangian in radial coordinates.  Under the decomposition
$M= \Omega \Lambda \Omega^\dagger$, the Lagrangian becomes
\be
{\cal L} = \sum_{i=1}^N \left ( \half \dot \lambda_i^2 - V(\lambda_i) \right ) +
\half \sum_{i \neq j} (\lambda_i - \lambda_j)^2 
\left ( \left (\dot \Omega \Omega^\dagger \right )_{ij} \right)^2~,
\ee
where $\dot \Omega \Omega^\dagger = -\dot \Omega^\dagger \Omega$
is traceless and anti-hermitian.
The classical Hamiltonian can be defined using canonical momenta
$P^\lambda_i$ and $P^{\Omega}_{ji}$
conjugate to $\lambda_i$ and $\Omega_{ij}$ respectively, and
equals
\be
H =  \sum_i \left( \half\left(P^\lambda_i  \right)^2 +  V(\lambda_i)\right)+
\half \sum_{i\neq j} {
\Pi_{ij} \Pi_{ji}
\over
(\lambda_i - \lambda_j)^2
}~,
\label{eqn:Hclass}
\ee
where $\Pi = \Omega(P^\Omega)^T$ are the generators of $SU(N)_R$,
the action of $SU(N)$ by multiplication of $\Omega$ on the right.
Notice that this is not the same as the gauge symmetry of the
Hamiltonian, which is $SU(N)_L$.

To understand the meaning and the action of $\Pi_{ij}$, let us repeat the
derivation of the Hamiltonian (\ref{eqn:Hclass}) in a small region
around a particular point in angular space, $\Omega_0$.  
Write $\Omega(t) = \Omega_0 \exp(iA(t))$, where we will assume that the
hermitian matrix $A$ is small, and ignore higher order terms in $A$.
Then, $\dot \Omega^\dagger \Omega = \dot A$ and the Lagrangian can
be written as
\be
{\cal L} =  \sum_{i=1}^N \left (\half\dot \lambda_i^2  - V(\lambda_i) \right )+
\half \sum_{i \neq j} (\lambda_i - \lambda_j)^2 
\left ( \dot A _{ij} \right)^2~.
\ee
It now becomes clear that $\Pi_{ij}$ is the conjugate momentum to
$A$, $\left [ A_{ij}, \Pi_{kl} \right ] = i \delta_{il} \delta_{jk}$,
or 
\be
\Pi_{ij} = -i {\partial \over \partial A_{ji}}~.
\ee
The action of $\Pi$ on $\Omega$ is therefore
\be
\Pi_{ij} \Omega_{kl} = \left .  -i {\partial \over \partial A_{ji}} 
\left( \Omega e^{iA} \right)_{kl} \right |_{A=0} = 
 \delta_{il}\Omega_{kj}
\ee
and, similarly,
\be
\Pi_{ij} \Omega^\dagger_{kl} = 
- \delta_{jk}\Omega^\dagger_{il} ~.
\ee
With these formulas on hand, we can use the Hamiltonian in
(\ref{eqn:Hclass}) to derive the Schrodinger equation in
any representation, as long as we take into account the
Jacobian for our change of coordinates $M \rightarrow
\Omega, \Lambda$.  This Jacobian is simply the square of the
Vandermonde determinant, $\Delta^2$, 
where $\Delta(\lambda) = \Pi_{i<j}(\lambda_i-\lambda_j)$.
We must rescale the wavefunction by $\Delta$;
therefore, when acting on wavefunctions of the form
\be
\tilde \Psi_\a(\Lambda,\Omega) = \Delta(\lambda)
\sum_{k=1}^K f^{(k)} (\lambda_1, \lambda_2,..,\lambda_N) \Omega_{\a\b} v^{(k)}_\b~,
\ee
the Hamiltonian is
\be
H = \sum_i \left(-\half {\partial^2 \over \partial \lambda_i^2}
 + V(\lambda_i) \right) +
\half \sum_{i\neq j} {
\Pi_{ij} \Pi_{ji}
\over
(\lambda_i - \lambda_j)^2
}~.
\label{eqn:Hradial}
\ee

\subsection{The singlet representation}

The simplest representation to work with is, of course,
the singlet, and this has been much studied.
Here, we will just restate a few of the salient facts in 
our notation.
We have $\Omega_{\a\b} = 1$ and therefore
\be
\Psi_{singlet}(M) = f_{symm}(\lambda_i)~.
\ee
and
\be
\tilde \Psi_{singlet}(M) =\Delta(\lambda) f_{symm}(\lambda_i) \equiv
f_{asym}(\lambda_i)
~.
\ee
The Hamiltonian acts as
\be
H f_{asym}(\lambda_i) = \sum_{i=1}^{N} \left [
-\half 
\left ( {\partial \over \partial \lambda_i} \right )^2
 + V(\lambda_i) \right ] f_{asym}(\lambda_i)~,
\ee
Since the wavefunction $f_{asym}(\lambda_i)$ is completely
asymmetric under permutations of $\lambda$s, 
this result is the the well known result that the problem
reduces to $N$ noninteracting fermions in the potential $V$,
where each fermion coordinate corresponds to a single eigenvalue.

\subsection{The adjoint}

This is the best studied non-singlet representation 
\cite{Marchesini:1979yq}, and the focus of the current paper.  
There are N-1 zero-weight vectors and
the problem requires solving N-1 coupled differential equations.

Wavefunctions in the adjoint are of the form
\be
\Psi_{ij} = \sum_{k=1}^{N} f_k(\lambda_i) \Omega_{ik} \Omega^\dagger_{kj}~,
\label{eqn:adj.wfn}
\ee
where we must impose a constraint 
\be
\sum_{k=1}^{N} f_k(\lambda_i) = 0~,
\label{adj-constraint}
\ee
while condition (\ref{eqn:permutations}) implies that
\be
f_{\sigma(k)}(\lambda_i) = f_{k}(\lambda_{\sigma(i)})~.
\label{eqn:AdjPermute}
\ee
There is no minus sign, since the phases acquired by $\Omega$ and 
$\Omega^\dagger$ cancel.
An arbitrary wavefunction in the adjoint representation,
including the Jacobian factor, is then
\be
\tilde \Psi_{adjoint} = \Delta(\lambda)
\sum_{k=1}^{N} f_k(\lambda_i) \left ( \Omega^\dagger C \Omega \right )_{kk}
\label{eqn:adj.wfn.2}
\ee
where $C$ is a traceless matrix of complex coefficients.

Notice that equation (\ref{eqn:AdjPermute}) allows us to find
any $f_k$ once one of them, say $f_1$, is known.  
$f_1$ is completely symmetric in $N-1$ eigenvalues $\lambda_2,
$\ldots,$\lambda_N$, with $\lambda_1$ `special'.  The
presence of this special eigenvalue leads us to
the impurity interpretation.

It is easy to check that 
\be
\Pi_{ij}\Pi_{ji}~\left ( \Omega^\dagger C \Omega \right )_{kk}
= \left ( \Omega^\dagger C \Omega \right )_{kk} 
\left ( \delta_{ik} + \delta_{jk}\right )
- \left ( \Omega^\dagger C \Omega \right )_{ii} \delta_{jk}
-\left ( \Omega^\dagger C \Omega \right )_{jj} \delta_{ik}
\ee
(with no summation implied anywhere in the above equation), 
and therefore
we obtain a set of Schrodinger equations for wavefunctions
$\tilde f_k(\lambda) = \Delta(\lambda_i) f_k(\lambda_i)$
\be
\sum_{i=1}^{N} \left [
-\half 
\left ( {\partial \over \partial \lambda_i} \right )^2
+ V(\lambda_i) \right ] 
\tilde f_k(\lambda_i) + \sum_{k' \neq k} 
{\tilde f_k(\lambda_i) - \tilde f_{k'}(\lambda_i) 
\over (\lambda_k - \lambda_{k'})^2}
= E \tilde f_k(\lambda_i)~.
\label{eqn:schrodinger-adj}
\ee
The double sum term can be thought of as an `impurity-hopping' 
interaction term.


\section{Collective field analysis}
\label{s2}

The main goal of this section is to extend the analysis in
\cite{Donos:2005vm} to include the backreaction from the 
impurity on eigenvalue density.
To this end, we will derive the exact (large N) 
collective field Hamiltonian in the adjoint sector.

Define
\bear
\psi(x,0) = \int {dk \over 2 \pi} e^{-ikx} \tr\left(e^{ikM}\right)~,
\\
\psi(x,1) = \int {dk \over 2 \pi} e^{-ikx} \tr\left(C e^{ikM}\right)~,
\eear
where $C$ is a traceless hermitian matrix.  
Notice that $\psi(x,0)$ is nothing more than the eigenvalue density
\be
\psi(x,0) = \sum_i\delta(x-\lambda_i) = \rho(x)~.
\ee
Wavefunction
(\ref{eqn:adj.wfn}) can be written as
\be
\Psi = \tr (C \Psi) = \int dx ~\psi(x,1) ~\Phi\left(x,\psi(\cdot,0)\right)~,
\label{eqn:adj.wfn.coll}
\ee
where $\Phi$ is a function (or, more correctly, a functional) of 
the variable $x$ and the field $\psi(\cdot,0)$.  In the adjoint sector,
then, we have a theory coupling the density of eigenvalues to
a single impurity at $x$.

In \cite{Donos:2005vm} the kinetic term of the Hamiltonian
(\ref{eqn:H}) is rewritten in terms of derivatives with respect
to the fields $\psi(\cdot,0)$ and $\psi(\cdot,1)$.  The result,
when acting on the wavefunction (\ref{eqn:adj.wfn.coll}), is
\bear
-{1 \over 2} {\partial \over \partial M_{ab}} 
{\partial \over \partial M_{ba}} =
 &-&{1 \over 2} \sum_{s=0,1} \int dx ~ \omega(x,s) 
{\partial \over \partial \psi(x,s)} \\ \nn
&-& {1 \over 2} \sum_{s,s'=0,1} \int dx \int dy~ \Omega(x,s;y,s')
{\partial^2 \over \partial \psi(x,s) \partial \psi(y,s')} ~,
\eear
where 
\bear
\omega(x,s) &=& 
{\partial^2 \psi(x,s) \over \partial M_{ab} \partial M_{ba}}~,
\\
\Omega(x,s;y,s') &=& {\partial \psi(x,s) \over \partial M_{ab}}
{\partial \psi(y,s') \over \partial M_{ba}}  ~.
\eear
Expressions for $\Omega$ and $\omega$ in terms of the collective fields
can be computed explicitly, either by using the formulas for the laplacian
acting on functions in the singlet and adjoint representations, 
or in Fourier space.  In the latter method, we use this simple identity
\be
{\partial \over \partial M_{ba}} 
\left ( e^{ikM} \right )_{cd} 
= ik \int_0^1 d\alpha~ \left ( e^{ik\alpha M} \right )_{cb} 
\left ( e^{ik(1-\alpha)M} \right )_{ad} ~.
\ee
The results are
\bear
\omega(x,0) &=& -2 \partial_x \left [ \psi(x,0) \int dy
{\psi(y,0) \over (x-y)}
  \right ] \label{o0}
\\
\omega(x,1) &=& -2 \partial_x \left [ \psi(x,1) \int dy
{\psi(y,0) \over (x-y)} \right ] \nn\\ &~& -~2 \left [
\psi(x,1) \int dy {\psi(y,0) \over (x-y)^2 }
-
\psi(x,0) \int dy {\psi(y,1) \over (x-y)^2 }
  \right ] \label{o1}
\\
\Omega(x,0;y,0) &=& \partial_x \partial_y \left(\psi((x+y)/2,0) 
\delta(x-y)\right)
\label{O0}
\\
\Omega(x,0;y,1) &=& \partial_x \partial_y \left(\psi((x+y)/2,1) 
\delta(x-y)\right)
\label{O1}
\eear
Notice that we will not need $\Omega(x,1;y,1)$, since the wavefunction
(\ref{eqn:adj.wfn.coll}) is linear in $\psi(x,1)$. 
(This is a good thing, since $\Omega(x,1;y,1)$ involves the collective
field for two impurities.)

All the integrals in equations (\ref{o0}) and (\ref{o1}) should be regularized
using the Principal Value  prescription, which arises when we
perform the following exchange of order of integration
\bear
\int_0^1 d\alpha~ \int dy~   (ik) e^{i\alpha k(x-y)} G(x,y) &=&
P.V. \int dy~ \int_0^1 d\alpha~ (ik) e^{i\alpha k(x-y)} G(x,y) \nn \\&=&
P.V. \int dy~ {e^{ik(x-y)} G(x,y) \over x-y}
\eear
where $G$ is some function.  The Principal Value prescription will
be implicit from now on in this section.

As was already seen in the previous section, the Jacobian $J$ for changing 
variables from the matrix to its eigenvalues is simply the Vandermonde 
determinant squared, $(\Delta(\lambda))^2$.  
Up to a sign, and ignoring a set of measure zero where
the eigenvalues coincide, we have
\be
\ln(J) = \sum_{i\neq j} \ln ( \lambda_i - \lambda_j) = 
\int dx dy~ \psi(x,0) \psi(y,0) \ln(x-y)~.
\ee

In order to arrive at a self-adjoint Hamiltonian, we must perform a similarity
transformation, equivalent to rescaling the wavefunction by the square root 
of $J$, which leads to the following substitution
\cite{Donos:2005vm}
\be
{\partial \over \partial \psi(x,0) } ~~ \rightarrow~~
{\partial \over \partial \psi(x,0) } - {1 \over 2}
{\partial \ln J \over \partial \psi(x,0) }~ = 
{\partial \over \partial \psi(x,0) } -  \int dy {\psi(y,0) \over x-y}~.
\ee

Finally, introducing the canonical conjugate to $\psi(x,0)$, 
\be
\Pi(x) = -i {\partial \over \partial \psi(x,0)}~.
\ee
and putting all of this together, we get that those terms in the Hamiltonian 
which involve only $s=0$ are
\bear
H_{s=0} &=& 
- \half \int dx \omega(x,0)  
\left ( {\partial \over \partial \psi(x,0) } - {1 \over 2}
{\partial \ln J \over \partial \psi(x,0) } \right )
\nn \\ &-&   \half 
 \int dx dy~ \Omega(x,0;y,0)
\left ( {\partial \over \partial \psi(x,0) } - {1 \over 2}
{\partial \ln J \over \partial \psi(x,0) } \right )
\left ( {\partial \over \partial \psi(y,0) } - {1 \over 2}
{\partial \ln J \over \partial \psi(y,0) } \right )
\nn \\ &+& 
\int dx  \psi(x,0) \left ( V(x) + \mu  \right)~,
\eear
which combine to
give the well-known collective field Hamiltonian
\bear
\label{Hs0}
H_{s=0} &=& \half \int dx ~ \partial_x \Pi(x) \psi(x,0) \partial_x \Pi(x)\
\\ \nn
&+&\int dx~ \left ( 
{\pi^2 \over 6} \psi^3(x,0) + \psi(x,0) \left (  V(x)+\mu\right)
\right )~
\\ \nn
 &+& \half \int dx \psi(x,0) \left ( 
\partial_x \partial_y \ln|y-x|_{y=x}
\right )~.
\eear

We have added a Lagrange multiplier $\mu$ to fix the total number of
eigenvalues.
To obtain the second line in (\ref{Hs0}), it is necessary to use that
\be
\int dx g(x)~ \left [ P.V.~ \int dy {g(y) \over x-y}  \right ]^2 =
{\pi^3 \over 3} \int dx (g(x))^3~,
\ee
which is easy to prove in Fourier space.

The last line is the 1-loop contribution to the energy \cite{Das:1990kaa}.

Terms containing  the $s=1$ fields contribute three more terms to
the Hamiltonian,
\bear
H_{s=1} &=& 
- \half \int dx \omega(x,1)  {\partial \over \partial \psi(x,1)} 
- \int dx dy~ \Omega(x,0;y,1)
{\partial^2 \over \partial \psi(x,0) \partial \psi(y,1)}
\nn \\ &+& \half 
\int dx dy~ \Omega(x,0;y,1)
{\partial \ln J \over  \partial \psi(x,0)}
{\partial \over  \partial \psi(y,1)}~,
\eear
which evaluate to
\bear
H_{s=1} &=& \int dx dy 
{\psi(x,1)  \psi(y,0) - \psi(x,0) \psi(y,1) \over (x-y)^2 }
  {\partial \over \partial \psi(x,1)} \nn \\ 
&-&
\int dx ~  \psi(x,1)
\left ( \partial_x {\partial \over \partial \psi(x,0)} \right )
 \left ( \partial_x {\partial \over \partial \psi(x,1)} \right ) ~, 
\eear
When acting with $H_{s=1}$ on the wavefunction (\ref{eqn:adj.wfn.coll}),
we get 
\bear
\label{Hs1}
& & H_{s=1} \Psi = \\ \nn & &
\int dx ~\psi(x,1) 
\left [ \int dy{ \psi(y,0) \over (x-y)^2}\left [ \Phi(x,\psi(\cdot,0))
- \Phi(y,\psi(\cdot,0)) \right ] +
 \partial_x \Pi(x) ~(-i\partial_x)  \Phi(x,\psi(\cdot,0))
\right ] 
\eear
which when added to the singlet sector Hamiltonian (\ref{Hs0})
gives the exact collective field Hamiltonian in the
adjoint sector.


\section{Free fermions}
\label{s3}

In this section, we will re-derive the Hamiltonian using second
quantized formalism for free fermions to represent the eigenvalues.

Let us define a function $F(x;\underline \lambda)$
of a coordinate $x$ and the set of eigenvalues $\underline \lambda$,  
completely antisymmetric in those eigenvalues.  We will need
this function to have the property that
\be
F(x,\underline \lambda) |_{x=\lambda_k} = 
\Delta(\underline \lambda) f_k(\lambda_i)~.
\ee
$F$ could  be defined as
\be
F(x,\underline \lambda) = \sum_k \left [
 \Delta(\underline \lambda)f_k(\lambda_i) \right ] |_{\lambda_k=x}
\ee
but it is not unique (see the next section for a discussion of this).

Another ingredient we need is two auxiliary fermion fields 
$\Psi_\uparrow(x)$ and $\Psi_\downarrow(x)$,
such that $\{\Psi_i(x),\Psi_j^\dagger(y)\} = \delta(x-y) \delta_{ij}$,
and a vacuum state $|{\bf 0}\rangle$ defined by 
$\Psi_i(x)|{\bf 0}\rangle = 0$.

Let us now define a state corresponding to the adjoint wavefunction
\be
|f^{adj}\rangle \equiv 
\left ( \prod_{i=1}^N \int d\lambda_i \right )
\sum_j ~\Delta(\underline\lambda)f_j(\lambda) ~
\Psi_\downarrow^\dagger(\lambda_1) \ldots\Psi_\downarrow^\dagger(\lambda_{j-1})
\Psi_\uparrow^\dagger(\lambda_j)
\Psi_\downarrow^\dagger(\lambda_{j+1}) \ldots\Psi_\downarrow^\dagger(\lambda_N)
|{\bf 0}\rangle
\ee
which can be rewritten in terms of $F$ as
\be
\label{fadj}
|f^{adj}\rangle \equiv 
\int dx \Psi_\uparrow^\dagger(x)\Psi_\downarrow(x)~~
\left ( \prod_{i=1}^N \int d\lambda_i \right )
F(x;\underline \lambda) 
\left ( \prod_{i=1}^N  \Psi_\downarrow^\dagger(\lambda_i)\right )
|{\bf 0}\rangle~.
\ee
The term in front, $\int dx \Psi_\uparrow^\dagger(x)\Psi_\downarrow(x)$,
is the same for all states, and can be stripped.  What we are left with
is
\be
\label{fermions-impurity}
\left ( \prod_{i=1}^N \int d\lambda_i \right )
F(x;\underline \lambda) 
\left ( \prod_{i=1}^N  \Psi_\downarrow^\dagger(\lambda_i)\right )
|{\bf 0}\rangle~,
\ee
which has the interpretation of a state in a theory of a single fermion field,
$\Psi_\downarrow$ and an impurity.  The $\uparrow$ fermions were only a crutch,
and disappear in this way of thinking about it.

To effect the stripping of the 
$\int dx \Psi_\uparrow^\dagger(x)\Psi_\downarrow(x)$ term,
we can multiply any state of the form (\ref{fadj}) on the left by
$\Psi_\downarrow^\dagger(z)\Psi_\uparrow(z)$.  This leads to
\be
\label{f-stripped}
\Psi_\downarrow^\dagger(z)\Psi_\uparrow(z) |f^{adj}\rangle =
\Psi_\downarrow^\dagger(z)\Psi_\downarrow(z)~
\left ( \prod_{i=1}^N \int d\lambda_i \right )
F(x;\underline \lambda) 
\left ( \prod_{i=1}^N  \Psi_\downarrow^\dagger(\lambda_i)\right )
|{\bf 0}\rangle~,
\ee
which is the same as equation (\ref{fermions-impurity}) except for 
a factor of $\Psi_\downarrow^\dagger(z)\Psi_\downarrow(z)$ in front,
which is nothing more than the fermion density at point $z$,
$\Psi_\downarrow^\dagger(z)\Psi_\downarrow(z) = \rho(z)
= \psi(z,0)$.

For ease of notation, let us define
\be
|\underline\lambda\rangle \equiv
\left ( \prod_{i=1}^N  \Psi_\downarrow^\dagger(\lambda_i)\right )
|{\bf 0}\rangle
\ee
and
\be
\int d \underline \lambda \equiv
\left ( \prod_{i=1}^N \int d\lambda_i \right )~.
\ee

The action of the Hamiltonian in this notation  can be derived by taking
the left hand side of equation (\ref{eqn:schrodinger-adj}), 
multiplying by $
\Psi_\downarrow^\dagger(\lambda_1) \ldots\Psi_\downarrow^\dagger(\lambda_{k-1})
\Psi_\uparrow^\dagger(\lambda_k)
\Psi_\downarrow^\dagger(\lambda_{k+1}) \ldots\Psi_\downarrow^\dagger(\lambda_N)
|{\bf 0}\rangle$, summing over $k$, integrating over all eigenvalues and 
rewriting everything in terms of $F$ instead of the $f_k$s.
Doing this carefully leads to a number of terms,
\bear
\label{Hf}
&& H |f^{adj}\rangle = \\ \nn
&& \int dx \Psi_\uparrow^\dagger(x)\Psi_\downarrow(x) ~~
\left [ -\half \int dy \Psi_\downarrow^\dagger(y) \partial_y^2
\Psi_\downarrow(y)  \right ]
\left (\int d \underline \lambda \right)
F(x;\underline \lambda)|\underline\lambda\rangle 
\\ \nn
&& + \int dx \Psi_\uparrow^\dagger(x)\Psi_\downarrow(x) ~~
\left [  \int dy \Psi_\downarrow^\dagger(y)
\left (  V(y) + \mu\right )
\Psi_\downarrow(y)  \right ]
\left ( \int d \underline \lambda \right )
F(x;\underline \lambda)|\underline\lambda\rangle  
\\ \nn 
&& + 
\half \int dx \left ( \partial_x\Psi_\uparrow^\dagger(x)\Psi_\downarrow(x)
-\Psi_\uparrow^\dagger(x)\partial_x \Psi_\downarrow(x) \right )
\left (\int d \underline \lambda \right )
\partial_x F(x;\underline \lambda) |\underline\lambda\rangle  
\\ \nn
&& + \int dx \Psi_\uparrow^\dagger(x)\Psi_\downarrow(x) ~~
\left ( \int d \underline \lambda \right )
P.V.\int dy {F(x;\underline\lambda)-F(y;\underline\lambda) \over (x-y)^2}
\Psi_\downarrow^\dagger(y)\Psi_\downarrow(y)
|\underline\lambda\rangle  ~.
\eear
The Principal Value appears because the sum in equation 
(\ref{eqn:schrodinger-adj}) excludes $k=k'$ terms.
It has the effect of allowing us to ignore the anticommutator
term between $\Psi_\downarrow(y)$ and $\Psi_\downarrow(x)$, since 
$x=y$ is not in the domain of integration. 
It also matches the P.V. prescription in the bosonic treatment.

The first two lines are nothing but the kinetic and potential
terms of a non-relativistic noninteracting fermion field Hamiltonian.
As is well known, after bosonization, these turn into the terms
in equation (\ref{Hs0}) \cite{Gross:1990st,Klebanov:1991qa}.

The last two lines should correspond to equation (\ref{Hs1}). 
The exchange term is clearly the same, but we need to look more
carefully at the third line of (\ref{Hf}) and the final term
in equation (\ref{Hs1}).  The difficulty here is that the 
universal term $\int dx \Psi_\uparrow^\dagger(x)\Psi_\downarrow(x)$
is not in front explicitly.  We will strip  it off by the method
which lead to equation (\ref{f-stripped}),
\bear
&& \Psi_\downarrow^\dagger(z)\Psi_\uparrow(z)~\times~
\half \int dx \left ( \partial_x\Psi_\uparrow^\dagger(x)\Psi_\downarrow(x)
-\Psi_\uparrow^\dagger(x)\partial_x \Psi_\downarrow(x) \right )
\left (\int d \underline \lambda \right )
\partial_x F(x;\underline \lambda) |\underline\lambda\rangle  \nn \\ 
&& =
\half  \left ( \partial_z\Psi_\downarrow^\dagger(z)\Psi_\downarrow(z)
-\Psi_\downarrow^\dagger(z)\partial_z \Psi_\downarrow(z) \right )
\left (\int d \underline \lambda \right )
\partial_x F(x;\underline \lambda) |\underline\lambda\rangle 
\eear
We recognize the multiplier in front as simply $-i$ times the total
momentum density of the fermions at point $z$, which will we denote 
with $-i J(z)$.  Comparing with equation (\ref{f-stripped}), the third
line of the Hamiltonian should be interpreted as
\be
-i ~ {J(z) \over \rho(z)} ~\partial_z~.
\ee
We wish to show that this is the same as $\partial_z \Pi(z) ~(-i\partial_z)$
in equation (\ref{Hs1}). 

From the Hamiltonian in equation (\ref{Hs0}), we have that
\be
{\partial \over \partial t} \rho(x) = -\partial_x \left(\rho(x) \partial_{x}
\Pi(x)\right)
\ee
The continuity equation implies that
\be
{\partial \over \partial t} \rho(x) = -\partial_x J(x) 
\ee 
and therefore
\be
J(x) = \rho(x) \partial_{x} \Pi(x)~,
\label{bosonisation-result}
\ee
as required.

Notice that in deriving the above correspondence we used
the bosonic equation of motion arising from the s=0 sector alone,
and the continuity equation for fermions which is implied by
the kinetic term in the first line of (\ref{Hf}).  In other words,
our result is the consequence of bosonization in the singlet sector.
We could have also used the bosonic equation of motion modified by
the terms is the s=1 sector, together with a new continuity
equation applicable to a fermionic field whose kinetic term is
affected by the last line of (\ref{Hf}), but this would have
been unnecessarily complicated, since
equation (\ref{bosonisation-result}) is simply an algebraic
consequence of bosonization itself and does not depend on the 
details of the dynamics.

\section{Discussion - collective field approach}
\label{s4}

We have now derived the collective field Hamiltonian which
describes the interaction of the impurity with the eigenvalue
density in two different ways.

It is worth mentioning here that the adjoint wavefunction 
$\Phi(x,\psi(\cdot,0))$, and its antisymmetric equivalent
$F(x,\underline \lambda)$, are not unique.  For example,
we can add any function of the form $\prod_k (x-\lambda_k) G$ 
to $F$, as long as $G(x,\underline \lambda)$ is regular
and antisymmetric in the $\lambda$s.  This ambiguity should
not play a role in the large $N$ limit.

The wavefunction is also subject to the adjoint constraint
(\ref{adj-constraint}),
\be
\int dx \psi(x,0) F(x;\underline\lambda) = 0~.
\ee

The ambiguity and the constraint would have to be taken into account 
when solving the equations of motion.  

As usual with the  c=1 matrix model, equations of motion should be 
solved perturbatively, in the semi-classical expansion,
with $N$ strictly infinite.  With the upside down harmonic
oscillator potential $V(x) = - \half x^2$, we can treat the
semi-classical expansion as an expansion in $\mu^{-1}$.
On the string theory side, this corresponds to expansion in
the string coupling, since $\mu$ = $g_{s}^{-1}$ \cite{Klebanov:1991qa}.
 Weak string coupling corresponds to
large $\mu$, and this is the limit we want to work with.

In order to better exhibit the dependence on $\mu$, we will rescale
the coordinate $x$ to $\sqrt \mu x$.  To maintain the shape of Fermi sea
in phase space, we must then rescale  
\be
\psi(\cdot,0) \rightarrow \sqrt \mu \psi(\cdot,0)~.
\ee
The scaling of $\Pi$ is determined from fixing
\be
\int dx {\partial \over \partial \psi(x,0)} \psi(y,0) = 1~,
\ee
to be
\be
\Pi(x) \equiv -i {\partial \over \partial \psi(x,0)}  \rightarrow \mu^{-1} 
\Pi(x)~.
\ee

The scaling of $\psi(x,1)$ does not matter, as it cancels out in 
the Hamiltonian.  Canonical commutation relationship require that
the combination $\psi(x,1)\partial/\partial\psi(x,1)$ scale as
\be
\psi(x,1){\partial \over \partial\psi(x,1)} \rightarrow
\mu^{-\half} \psi(x,1){\partial \over \partial\psi(x,1)}
\ee

With this scaling, the functions in (\ref{o0})-(\ref{O1}) 
scale as
\bear
\omega(x,s) &\rightarrow& \sqrt \mu \omega(x,s)~, \nn \\
\Omega(x,0,y,1) &\rightarrow& \mu^{-1} \Omega(x,0,y,1)~,  \\
\Omega(x,0,y,1) {\partial \over \partial \psi(y,1) }
&\rightarrow& \mu^{-2} 
\Omega(x,0,y,1) {\partial \over \partial \psi(y,1) }~.
\nn
\eear

After this rescaling, the leading term in the Hamiltonian
is
\be
\mu^2 \int dx \left ({\pi^2 \over 6} \psi^3(x,0) + 
(\mu - x^2/2)\psi(x,0)
  \right)~,
\ee
which comes from the $s=0$ part of the Hamiltonian.
The classical solution is  the
well-known eigenvalue density distribution given by
\be
\pi \psi(x,0) = \pi \phi_0(x) = \sqrt{x^2 - 2}~,
\ee
where $\mu$ is the Fermi level energy.  Allowing the density
to fluctuate,
\be
 \psi(x,0) =  \phi_0(x)+ {1 \over \sqrt \pi \mu} \partial_x \eta~,
\ee
we can obtain the $o(\mu^0)$ term in the large $\mu$ expansion of
the Hamiltonian.  Let us focus first on that part of the Hamiltonian
which contains only $s=0$ field.  The desired term is
\be
\half \int dx \pi \phi_0(x) \left( P^2 + 
(\partial_x \eta)^2 \right ) 
\ee
where $P = -\partial_x \Pi(x) /  (\sqrt \pi \mu)$ 
is the conjugate momentum to $\eta$.
Notice that the $\mu$-scaling of $\eta$ is necessary to obtain
this (quite standard) massless scalar kinetic term.
Interactions are higher order in $\mu^{-1}$:
\be
\mu^{-1} \int dx \sqrt \pi \left ({1 \over 6} (\partial_x \eta)^3
+ \half P^2 \partial_x \eta \right )~.
\ee

Now, let us turn our attention to the terms involving the
$s=1$ field.  There are three such terms in the Hamiltonian,
and their $\mu$-scalings are
\bear
\int dx \omega(x,1)  {\partial \over \partial \psi(y,1)} 
&\rightarrow&
\mu^0~\int dx \omega(x,1)  {\partial \over \partial \psi(y,1)}
~, \\
\int dx dy~ \Omega(x,0;y,1)
{\partial^2 \over \partial \psi(x,0) \partial \psi(y,1)}
&\rightarrow&
\mu^{-2}
\int dx dy~ \Omega(x,0;y,1)
{\partial^2 \over \partial \psi(x,0) \partial \psi(y,1)}
~, \nn \\
\int dx dy~ \Omega(x,0;y,1)
{\partial \ln J \over  \partial \psi(x,0)}
{\partial \over  \partial \psi(y,1)}
&\rightarrow&
\mu^{0}
\int dx dy~ \Omega(x,0;y,1)
{\partial \ln J \over  \partial \psi(y,0)}
{\partial \over  \partial \psi(y,1)}
~. \nn
\eear
The first and the third terms are linear in $\psi(x,0)$.
Any term containing $\partial_x\eta$ as opposed to $\phi_0$
is therefore suppressed by $1/\mu$.  
The second term is linear in $\Pi \sim \mu P$, so any term containing
$P$ is also suppressed by $1/\mu$.
All the terms of order $\mu^0$ are thus those collected in
\cite{Donos:2005vm} equation (14).  Therefore, order $\mu^0$
part of the Hamiltonian involving $s=1$ fields is \cite{Donos:2005vm}
\be
\int dx dy 
{\phi_0(y) \psi(x,1) - \phi_0(x) \psi(y,1)
\over (x-y)^2
} {\partial \over \partial \psi(x,1)}~.
\ee
Notice that this is the Hamiltonian stated in \cite{Maldacena:2005hi},
whose eigenfunctions were found in \cite{Fidkowski:2005ck}.

To next order, as we have seen above, we have self-interactions of $\eta$, 
as well as coupling between $\eta$ and the position of the impurity, 
which is are the terms we are interested in,
\bear
&\mu^{-1}& \int dx dy {1\over \sqrt \pi}
{\partial_y \eta(y) \psi(x,1) - \partial_x \eta(x) \psi(y,1)
\over (x-y)^2
} {\partial \over \partial \psi(x,1)} \nn \\
&-&
\mu^{-1}~ \int dx ~\sqrt \pi ~ P(x) ~\psi(x,1)
 \left (-i \partial_x {\partial \over \partial \psi(x,1)} \right ) ~.
\label{impurity-interaction}
\eear

Interactions between the impurity and the collective field are of the 
same order as three-vertex self-interactions of the collective field.
This corresponds to the statement that the three point function of 
three tachyons on a sphere is of the same order in $g_{s}$ 
($g_{s}^1$) as the disk amplitude with two boundary and one bulk 
tachyon insertions.

\section{Scattering in the chiral formalism}
\label{s5}

While the Hamiltonian of the previous sections is quite enlightening,
it is not the best starting point  for a calculation of scattering amplitudes.
As was pointed out in \cite{Kostov:2006dy}, those are much easier to calculate 
in the chiral formalism \cite{Alexandrov:2002fh}.  We will extend the formalism in
\cite{Kostov:2006dy} to include backreaction on the collective field.

Methods presented so far were applicable to any potential.
The chiral formalism we are about to introduce is applicable only to
the $V=-\half x^2$ potential, which is the one relevant for c=1 string theory.
The chiral formalism for the up-side-down harmonic oscillator 
is similar to the raising and lowering operator formalism for the 
ordinary harmonic oscillator.  We will make a canonical transformation to
the light-cone (or chiral) variables
\be
M_\pm \equiv { M \pm P_M \over \sqrt 2 }~,
\ee
where $PM$ is the canonical conjugate to the matrix $M$.
The Hamiltonian in these variables is
\be
H = -{\half}Tr \left( M_+ M_- + M_- M_+\right )~.
\ee
Since equations of motion imply that $M_\pm$ evolves like
$e^{\pm t}$, in the far past we have $M \sim M_+$ and in the
far future $M \sim M_-$.  Therefore, wavefunctions written in
$M_+$ coordinates correspond to incoming states and written
in $M_-$ coordinates correspond to outgoing states.  Since
$M_+$ and $M_-$ are complimentary, these wavefunctions are
related by a Fourier transform, and the scattering
matrix is given by an inner product
\be
\langle \Psi^{-} | \Psi^{+} \rangle = 
\int dM_+  dM_- ~e^{i \tr(M_+ M_-)} ~
\sum_{\alpha=1}^{dim {\cal R}} ~
\overline {\Psi_\alpha^{-}(M_-)}~ \Psi_\alpha^{+}(M_+)~.
\ee
From our formalism in Section \ref{s1}, it is clear that the
integral $\int dM_+  dM_-$ can be performed separately over the angular
degrees of freedom $\Omega_\pm$ and the eigenvalue degrees of freedom
$\Lambda_\pm$.  As before, we have
$M_\pm = \Omega_\pm \Lambda_\pm \Omega_\pm^\dagger$ and
$dM_\pm = \prod_i \lambda_i^\pm \Delta(\lambda^\pm) d\Omega_\pm$.
Therefore
\be
\langle \Psi^{-} | \Psi^{+} \rangle = 
\int d\lambda^+_i d\lambda^-_j ~\Delta^2(\lambda^+) \Delta^2(\lambda^-)
~d\Omega~  \overline {\Psi_\alpha^{-}(\Lambda_-)}~ 
\Omega_{\alpha\beta} ~ \Psi_\beta^{+}(\Lambda_+)
~ e^{i \tr( \Lambda_+ \Omega \Lambda_- \Omega^\dagger)}~,
\ee
where $\Omega = \Omega_+^\dagger \Omega_-$ and 
$\Lambda_\pm = \mathrm{diag}(\lambda_1^\pm, \ldots, \lambda_N^\pm)$. 
The $\Omega$ integral can be evaluated with help from
the Morozov-Eynard formula \cite{Eynard:2004pz}
\bear
\label{mor-eyn}
\int d\Omega ~\tr \left ( {1 \over \xi_- + \Lambda_-} \Omega 
{1 \over \xi_+ + \Lambda_+} \Omega^\dagger  \right )
~ e^{i \tr( \Lambda_+ \Omega \Lambda_- \Omega^\dagger)} \\ \nn
= {\det \left (S + i {1 \over \xi_- + \Lambda_-} S 
{1 \over \xi_+ + \Lambda_+} \right ) - \det S \over 
 i \Delta(\lambda^+) \Delta(\lambda^-)}~,
\eear
where $\xi_\pm$ are arbitrary numbers and $S$ is a matrix with
elements $S_{jk} = \exp(i \lambda^+_j \lambda^-_k)$.

In a now familiar approach, we will absorb a factor of the Vandermonde
determinant into the wavefunction
\be
\label{wfpm}
\tilde \Psi^\pm (M_\pm) = \Delta(\lambda_\pm) \Psi^\pm (M_\pm)
\ee
as well as write a complete set of wavefunctions in the adjoint sector as
\be
\label{adj-chiral-basis}
\tilde \Psi^\pm (M_\pm) = \left ( 1 \over \xi_\pm + M_\pm \right )
\det_{kl} \left [ \psi^\pm_{E^\pm_k} (\lambda^\pm_l) \right ]~,
\ee
where the final factor is a Slater determinant of one-fermion eigenfunctions.
In the chiral formalism, the Hamiltonian acts on the wavefunctions
(\ref{wfpm}) as
\be
H = \mp i \sum_k \left ( \lambda_k^\pm {\partial \over \partial \lambda_k^\pm}
 + \half\right )
\ee
and  the one-fermion eigenfunctions are quite simple
\be
\psi^\pm(x^\pm) = {1\over \sqrt{2\pi}} ~x_\pm^{-\half \pm iE}~.
\ee

\begin{figure}
\ifpdf
\includegraphics[scale=1]{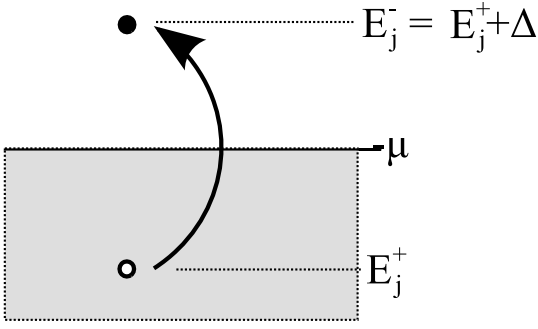}
\else
\includegraphics[scale=1]{fermion-up.eps}
\fi
\caption{Exciting one fermion above the Fermi sea level at $-\mu$.
\label{f2}}
\end{figure}

The first factor in equation (\ref{adj-chiral-basis}) is tailored
to the use of the Morozov-Eynard formula (\ref{mor-eyn}).  We can interpret
equation (\ref{adj-chiral-basis}) as representing a collective density of 
eigenvalues given by the Slater determinant together with an impurity
wavefunction given by $1/(\xi_\pm + x_\pm)$ for the incoming and 
outgoing states.  We can convert the impurity wavefunction basis
to the chiral eigenbasis with the use of this formula:
\be
\label{mellin}
\int_0^\infty d\xi ~ \xi^{iE} {1 \over \xi +  x} =
 {i \pi \over \sinh \pi E}~ x^{iE}~.
\ee
The inverse is also going to be useful,
\be
{i \over 2 \xi} \int dE {x^{iE}\xi^{-iE}\over \sinh(\pi E)} 
~=~{1\over \xi + x}~.
\ee
We wish to compute the in-out overlap
$\langle \Psi^- | \Psi^+ \rangle$ as a function of the
impurity `positions' $\xi_\pm$ and the occupied energy 
levels $\{ E^\pm_k\}$.  In \cite{Kostov:2006dy}, this was computed in
the case where  $E^+_k =  E^-_k = E_k$, where the energy levels
$E_k$ fill the potential up to level $\mu$ below the top.
That computation gave the scattering phase for the impurity 
while keeping the Fermi sea in its ground state.
We are interested in finding
the simplest correction to this scattering involving a change
in the Fermi sea.  We will therefore take $E^-_j =  E^+_j + \Delta$
for a single fermion labeled by $j$, 
and $E^+_k =  E^-_k$ for all $k\neq j$.  Choosing to study energy
emission from the long string, we take $\Delta >0$.  
This will correspond to a single fermion jumping up
from the Fermi sea and will be non-zero only if the incoming energy
$E^+_j$ is below the Fermi energy  $E^+_j<-\mu$ and the outgoing
energy $E^-_j = E^+_j + \Delta$
is above the Fermi sea level, $E^+_j + \Delta > -\mu$ 
Note that since we are taking $\mu$ very large,
$E_j^- < 0$, so that nothing spills to the other side of the 
potential. See Figure \ref{f2} for a graphical representation.

Let us make a few definitions, following Kostov \cite{Kostov:2006dy}.  
The single fermion reflection factor is defined via
\be
\int_0^\infty dx_+ dx_-~  \overline{\psi^-_E(x_-)} \psi^+_{E'}(x_+)
e^{ix_+ x_-}= \delta(E-E')   R(E) 
\ee
and is equal to\footnote{
The `phase factor' in equation (\ref{RE}) is not exactly unitary;
there exists a small overlap with the wavefunctions on the other
side of the potential.  Since we are working in the semi-classical
approximation, we can ignore this.}
\be
R(E) = \int_0^\infty {dr \over \sqrt r} r^{iE} e^{ir} = 
e^{i\pi/4} e^{-{\pi\over 2} E } \Gamma \left (\half + iE\right)~.
\label{RE}
\ee
We also define
\be
K(E',E) \equiv { 
\int_0^\infty dx_+ dx_- ~ \overline{\psi^-_E(x_-)} \psi^+_{E'}(x_+)~
{e^{ix_+ x_-}~ \over (\xi_+ + x_+ )(\xi_- + x_-)}
\over
\sqrt{R(E) R(E')}}~,
\ee
which can be evaluated  to give \cite{Kostov:2006dy}
\be
\label{eqn:K}
K(-E+\epsilon, -E-\epsilon) = 
{\left ( \xi_+ ^2 \over E\right )^{i\epsilon} - 
\left (  \xi_-^2 \over E\right )^{-i\epsilon}
\over i \sinh(2\pi\epsilon)}
{\pi \over \xi_+ \xi_- - E}~,
\ee
where $\epsilon \ll E$ and $E$ is positive.

Using the Morozov-Eynard formula, Kostov in \cite{Kostov:2006dy}
shows that any amplitude can be written as a determinant
\bear
&& \langle \Psi^-, \xi_-, E_j^- | \Psi^+ ,\xi_+, E_j^+  \rangle =
\\ \nn
&& -i \det_{nm}\left ( 
\delta(E_n^- - E_m^+) + iK(E_n^-, E_m^+)
\right )
\sqrt{\prod_{k=1}^N R(E^+_k) ~\prod_{k=1}^N R(E^-_k)}
\label{kostov-det}
\eear

In \cite{Kostov:2006dy}, the main result is that if there were no
fermions excited (so that $E_k^+ = E_k^- = E_k$ for all $k$) then,
\be
\langle \Psi^-, \xi_-| \Psi^+ ,\xi_+ \rangle =
\left ( \prod_k R(E_k) \right )~  e^{iS(\xi_+\xi_-)}~.
\ee
The first factor is simply the product of all the scattering
phases for all the fermions, an overall phase which is not
very interesting and which we will  ignore. 
 The second phase is for the scattering of the 
impurity, which depends (by time invariance) only on the combination
$\xi_+\xi_-$, and is related to the phase for scattering in an
energy eigenstate via
\be
e^{iS(\xi_+ \xi_-)} = - {1 \over 4 \xi_+ \xi_-}
 \int dE~ {(\xi_+ \xi_-)^{-iE} \over (\sinh (\pi E))^2 }
e^{-i \delta^{Adj}(E)}~.
\label{eqn:eS}
\ee
$\delta^{Adj}$ is the scattering phase for an impurity in an
energy eigenstate, and the explicit expression computed in
\cite{Kostov:2006dy} 
agrees with the value computed in string theory.

We are interested in a case where $E_k^+ = E_k^-$  for
all but one value of $k$.  The dominant contribution from the
determinant in (\ref{kostov-det}) comes from the term containing
the on-the-diagonal entry
$K(E_j^-,E_j^+)$ and a $(N-1)\times(N-1)$ determinant over the other
fermions for which it is true that  $E_k^+ = E_k^-$.  The latter factor
in the large $N$ limit is the same as the determinant leading to
the phase in (\ref{eqn:eS}).  Therefore, up to an overall
constant phase, which includes  $\prod_k R(E_k)$, we have
\be
\langle \Psi^-, \xi_-, E_j^- | \Psi^+ ,\xi_+, E_j^+  \rangle =
K(E_j^-,E_j^+) ~\sqrt{R(E_j^-)/R(E_j^+)} ~e^{iS(\xi_+ \xi_-)}~.
\label{overlap}
\ee

We will take the Mellin transform (\ref{mellin}) to express this
in terms of impurity incoming and outgoing energy eigenstates
with energies $E_{imp}^\pm$
\bear
&&\langle  E_{imp}^-, E_j^- | E_{imp}^+, E_j^+  \rangle = 
\\ \nn &&
- {\sinh (\pi  E_{imp}^-)\sinh (\pi  E_{imp}^+)   \over \pi^2}
\int_0^\infty d\xi_- ~ \xi_-^{iE_{imp}^-} ~
\int_0^\infty d\xi_+ ~ \xi_+^{iE_{imp}^+}
\langle \Psi^-, \xi_-, E_j^- | \Psi^+ ,\xi_+, E_j^+  \rangle ~.
\eear
We can now substitute  (\ref{eqn:eS}) and 
(\ref{overlap}) into the above, and obtain
\bear
&&\langle  E_{imp}^-, E_j^- | E_{imp}^+, E_j^+  \rangle = 
\\ \nn &&
{ \sinh (\pi  E_{imp}^-)\sinh (\pi  E_{imp}^+)   \over 4 \pi^2}
~\sqrt{R(E_j^-)/R(E_j^+)}
 \int dE~{e^{-i \delta^{Adj}(E)} \over (\sinh (\pi E))^2 } 
\\ \nn &&
~~\times ~\int_0^\infty d\xi_- ~ \xi_-^{iE_{imp}^-} ~
\int_0^\infty d\xi_+ ~ \xi_+^{iE_{imp}^+}~
(\xi_+ \xi_-)^{-iE-1}  K(E_j^+,E_j^+ + \Delta)~.
\eear
The double integral in the last line can be evaluated using
the explicit formula in equation (\ref{eqn:K}) and the substitution
$\xi_\pm = \sqrt \rho e^{\pm \sigma}$, 
\bear \nn
&& -{\pi \over i \sinh(\pi\Delta)} 
\int d\xi_- d\xi_+ ~ \xi_-^{i(E_{imp}^--E)-1} ~~ \xi_+^{i(E_{imp}^+-E)-1}~
{\left ( \xi_+^2  \over \tilde E\right )^{-i\Delta / 2} - 
  \left ( \xi_-^2 \over \tilde E\right )^{i\Delta /2 }
\over 
 \xi_+ \xi_- - \tilde E} 
 \\ && \nn 
=~{\pi \over i \sinh(\pi\Delta)}
\int_0^\infty {d\rho \over \rho}  \rho^{ik}
{\left ( \rho \over \tilde E \right )^{i\Delta/2} - 
\left ( \rho \over \tilde E \right )^{-i\Delta/2}
 \over \rho - \tilde E}
~
\int_{-\infty}^{+\infty} d\sigma e^{i(E_{imp}^+-E_{imp}^--\Delta)\sigma} 
\\ && =~
{4\pi^3 \over \sinh(\pi\Delta)} ~\delta (E_{imp}^+-\Delta-E_{imp}^-)
~~\tilde E^{ik-1} \Theta(\Delta/2 - |k|)~,
\eear
where $\tilde E \equiv -(E_j^+ + E_j^-)/2 >0$, and
$k =  (E_{imp}^++E_{imp}^-)/2 - E$.
Notice the  delta-function, which enforces energy conservation.
Plugging this back in,
\bear
\label{eqn:answerf}
&&\langle  E_{imp}^-, E_j^- | E_{imp}^+, E_j^+  \rangle = 
~\delta (E_{imp}^-+\Delta-E_{imp}^+)
\\ \nn &&
~~\times ~\pi~ \sinh (\pi  E_{imp}^-)\sinh (\pi  E_{imp}^+)   
~\sqrt{R(E_j^-)/R(E_j^+)}
\\ \nn && ~~\times ~
 \int_{E_{imp}^-}^{E_{imp}^+} dE~
{e^{-i \delta^{Adj}(E)} \over (\sinh (\pi E))^2 } 
\left( - {E_j^+ +E_j^- \over 2} \right)^{i \left (
(E_{imp}^+ +E_{imp}^-)/2 -E \right ) - 1}~.
\eear

To compare this answer with the string theory amplitude, the 
outgoing fermionic excitation has to be bosonized. This is 
relatively simple to do.  Start with the standard bosonization
formula for a relativistic fermion field $\Psi_R(\sigma)$
\be
:\Psi_R^\dagger(\sigma) \Psi_R(\sigma): ~= - {1 \over \sqrt \pi}
 \partial_\sigma \eta
\ee
and take a Fourier transform
\be
\alpha(\Delta) \equiv i \sqrt 2 \Delta  \int d\sigma
e^{i\sigma\Delta} \eta(\sigma) = 
\sqrt{2\pi} \int_{-\Delta/2}^{\Delta/2}{d \omega \over 2 \pi} 
:\Psi_R^\dagger(\omega+\Delta/2) \Psi_R(\omega-\Delta/2):~.
\label{alpha}
\ee
When acting on the fermion vacuum, the integrand is nonzero only if
 $\omega-\Delta/2 < 0$ and $\omega + \Delta/2> 0$, 
which is reflected in the choice of range of integration.
For $\Delta \neq 0$, we can drop the normal ordering.

It is a well known fact \cite{Klebanov:1991qa} that the fermionic eigenvalues
in the matrix model can be treated approximately as relativistic when
they are close to the Fermi surface.  The fermion creation and
destruction operators stay the same, but their interpretation changes:
for $\nu>0$
$\Psi_R(\nu)^\dagger$ used to create a fermion with energy $\nu$, it now
creates one with energy $-\mu+\nu$, and
$\Psi_R(-\nu)$ used to create an anti-fermion with energy $-\nu$, it now
creates a hole in the Fermi sea at energy $\mu-\nu$.  Therefore, writing
(\ref{alpha}) in terms of the non-relativistic fermions in Section \ref{s1} 
we get
\be
\alpha(\Delta) =
\sqrt{2\pi} \int_{-\Delta/2}^{\Delta/2}{d \omega \over 2 \pi} 
:\Psi(-\mu+\omega+\Delta/2)^\dagger \Psi(-\mu+\omega-\Delta/2):
\label{alpha2}
\ee
Notice that the fermion bilinear accomplishes exactly the emission
process shown in Figure \ref{f1}, with $E_j^+ = -\mu+\omega-\Delta/2$ and 
$E_j^- = -\mu+\omega+\Delta/2$.

Combining equation (\ref{alpha2}) 
with equation (\ref{eqn:answerf}), we get, up to an overall phase,
the amplitude for emission of a single quantum of the collective
field with energy $\Delta$ 
\bear
\label{eqn:answerb}
&& {\cal A}(E_{imp}^-,E_{imp}^+,\Delta) = 
\sqrt{\pi\over 2}~ \delta (E_{imp}^-+\Delta-E_{imp}^+)
 \sinh (\pi  E_{imp}^-)\sinh (\pi  E_{imp}^+)   
\nn \\  &&
~~\times ~\int_{-\Delta/2}^{\Delta/2} d\omega
~\sqrt{R(-\mu+\omega+\Delta/2) / R(-\mu+\omega-\Delta/2)}
\\ \nn &&
~~\times ~\int_{E_{imp}^-}^{E_{imp}^+} dE~
{e^{-i \delta^{Adj}(E)} \over (\sinh (\pi E))^2 } 
~\left( \mu - \omega \right)^{i \left (
(E_{imp}^+ +E_{imp}^-)/2 -E \right ) - 1}~.
\eear

\section{Conclusions and future directions}

Equations (\ref{Hs1}) (or (\ref{impurity-interaction})) 
and (\ref{eqn:answerb}) constitute
the main results of this paper.  (\ref{impurity-interaction}) describes
interactions between the tip of the long string and the background
collective field.  These interactions lead to a backreaction on the
collective field from the long string scattering.  The simplest 
amplitude induced by such interaction, where one quantum of the 
collective field is excited, is given in equation (\ref{eqn:answerb}).
As has already been mentioned in the introduction, it would
be interesting to compare this amplitude with the corresponding 
disk three-point function in boundary Liouville theory.
Agreement would strengthen the conjecture that long strings
correspond to the adjoint sector of the matrix quantum mechanics.

Further analysis of the amplitude in (\ref{eqn:answerb})
might help us understand the following puzzle.  In the matrix
model, the presence of a long string is associated with an 
impurity, which is a point-like object.  The long string itself,
however, is a space-filling object, and its interactions with the
tachyon field should reflect that somehow.  It would be interesting to
explore how these two opposing points of view are reconciled.

It would also be interesting to extend our analysis to multiple
quanta of tachyon excitations.  Potentially, this might lead to
a better understanding of the off-shell degrees of freedom in
spacetime physics (such as the graviton).  The long string should 
gravitate, and a scattering amplitude
involving an incoming and an outgoing tachyon might, for example,
contain a recognizable piece due to scattering from its gravitational field.
A related approach would be to try to solve equations of motion
implied by (\ref{Hs0}) and (\ref{Hs1}) together, using perturbation theory
outlined in section \ref{s4}.

\section*{Acknowledgments}

This work is supported by the Natural Sciences and Engineering
Research Council of Canada.  The author would like to acknowledge
useful comments from  Giovanni Felder, Matheson Longton and 
Andy Strominger.

\bibliographystyle{JHEP}
\bibliography{my}

\end{document}